\documentstyle[12pt,epsfig]{article}
\begin{document}
\begin{center}
\vspace{1.5in}
{\LARGE A double potential model for neutron halo nuclei }
\end{center}
\vspace{.4in}
\begin{center}
{\bf AFSAR ABBAS}\\
Institute of Physics\\ 
Bhubaneshwar-751005, India\\
\vspace{,1in}
email: afsar@iopb.res.in
\end{center}
\vspace{1.5in}
\begin{center}
{\bf Abstract}
\end{center}
\vspace{.3in}

It is shown here that loosely bound halo structure of neutron 
rich nuclei and the ground state spin of single neutron 
halo nuclei are correlated and are consistently explained
if one assumes a double potential shell model for these nuclei.

\newpage

Halo nuclei offer unique challenges in nuclear physics. It seems that
around a core of a few nucleons, there sit one or two nucleons very far 
away from the centre of the whole nucleus but still bound to it. 
As such their radii are much larger than what should be as per 
our understanding in standard nuclear physics [1]. Another feature 
is that for all the single neutron halo nuclei discovered so far the 
ground state spin has always been ${1 \over 2}^+$. At present this is 
being interprested as shifting and mixing  of single particle levels in 
the shell model [1,2]. Here we make a novel suggestion of a double 
potential shell model which accounts for both the loose halo structure 
and its ground state spin. Hence it is shown that the two are 
correlated in a basic way in this new model.

As of now a few two neutron halo nuclei and a few single neutron
halo nuclei have been discoverd. Howeevr what is puzzling is 
the experimenatl fact that for all the single neutron halo nuclei 
[1-3] ( which are   $^{11}_{4} Be_{7}$,  $^{15}_{6} C_{9}$  
and $^{19}_{6} C_{13}$ ) the 
ground state spin has always been  ${1 \over 2}^+$. 
At present this is being viewed
as shifting and mixing of different levels in these diverse nuclei 
in ways which may justfiably be called " mysterious and arbitaray ".
The author feels that for such unique and global features as
very loose structure of the valence neutron and the unique 
ground state spin of  ${1 \over 2}^+$ in all the cases known there 
has to be some common conncetion which should be basic and "simple".

What is it that these empirical features are hinting at.
Let us use the fact that the core is almost decoupled from the 
valence neutrons. In simple shell model we know the the 
central potential follows rather closely the nuclear density 
distribution [4]. 

\begin{equation}
V = - {V_0} \int_{0}^{\infty} \rho(r) {r^2} dr
\end{equation}

Let us put the empirical fact that the total density 
in a halo nucleus may be written as 
made up of a core and a halo part

\begin{equation}
\rho(r) = {\rho_{core}} + {\rho_{halo}}
\end{equation}

And thus the total potential separates out as

\begin{equation}
V = {V_{core}} + {V_{halo}}
\end{equation}

Hence the author here makes a suggestion that these features are
hinting at the existence of double potential for these nuclei.
A schematic plot of such a double potential well is given in Figure 1.
Let us take ${^{11}_{4} Be_{7}}$ as an example. Here as per empirical 
information  ${^{10}_{4} Be_{6}}$ forms a core and a single neutron orbiting 
it in a loose halo around this core. We assume tha the core 
${^{10}_{4} Be_{6}}$ is built as per standard shell structure in the inner
potential in Fig 1. The second potential structure assumes that it exists
becaues there exists a primary inner potential. Now the last valence 
neutron fills the orbits in the second potentail starting with the s-state. 
Hence its spin is  ${1 \over 2}^+$ which also is the spin of the whole 
nucleus ${^{11}_{4} Be_{7}}$. This is the situation in all the 
single neutron halo nuclei. The core
is built upon the inner potential as per standard nuclear shell model and 
the last valence neutron sits in the lowest orbital to give total ground 
state spin to the nucleus. If there are two neutrons in the halo then its 
apin contribution would be zero. Quite clearly in this picture if one 
were to add any more protons to halo nucleus it will go to the inner shell 
for these neutron rich nuclei. Quite clearly this double potential is 
for valence neutrons only. 
So for ${^{11}_{3} Li_{8}}$ as the 2- valence neutrons which form the halo 
would sit in the second potentail s-state and thus contribute spin zero.
Thus the total spin of the nucleus would come from the protons as per 
these filling up the first potetial shell.

This picture predicts that as the lowest orbital in the second potential 
can absorb only upto two neutrons so halos should most likely be 
observed for upto two neutrons. In all confirmed cases this is what the 
situation is, If one wants to add more neutron one has to send it to
the higher p- state. This would still be permitted but would require 
higher excitation energy. So higher number of neutrons in the halo is not 
ruled out as per this model. However cases with more than 2-neutrons in 
the halos will be suppressed.


\vspace{.2in}

{\bf References} 

\vspace{.2in}

1. I. Tanihata, Nucl. Phys. {\bf A682}, (2001) 114c

\vspace{.1in}

2. R. Kanungo, I. Tanihata and A. Ozawa, 
Phys. Lett.  {\bf B 528} (2002) 58

\vspace{.1in}

3. M. V. Zhukov, Eur. Phys. J. {\bf A 13} ( 2002) 27

\vspace{.1in}

4. P. E. Hodgson. Contemp. Phys., {\bf 35} (1994) 329

\newpage

\begin{figure}
\caption{Schematic double well potential to explain the properties 
of halo nuclei}
\epsfclipon
\epsfxsize=0.99\textwidth
\epsfbox{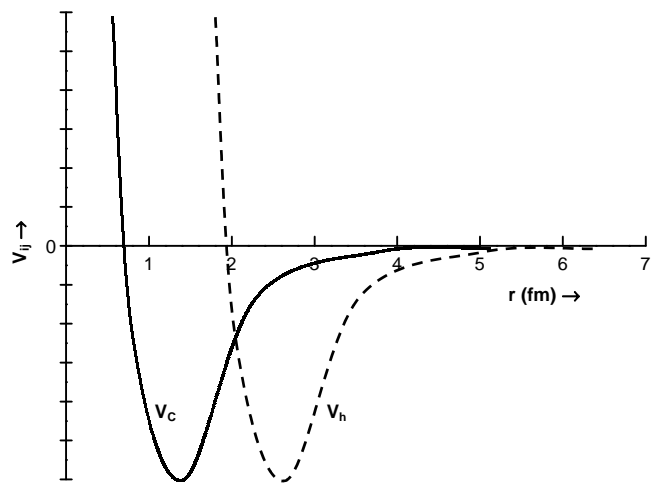}
\end{figure}

\end{document}